\RequirePackage{fix-cm}
\documentclass[smallextended]{svjour3}       % onecolumn (second format)
\smartqed  % flush right qed marks, e.g. at end of proof
\usepackage{graphicx}
\begin{document}

\title{Improvement of two-way continuous-variable quantum key distribution with virtual photon subtraction%\thanks{Grants or other notes
}

\titlerunning{Improvement of two-way continuous-variable quantum key distribution}        % if too long for running head

\author{Yijia Zhao         \and
        Yichen Zhang  \and
        Zhengyu Li    \and
        Song Yu      \and
        Hong Guo %etc.
}


\institute{Yijia Zhao \and Yichen Zhang \and Song Yu\at
              State Key Laboratory of Information Photonics and Optical Communications, Beijing University of Posts and Telecommunications, Beijing 100876, China \\
              Tel.: +86-010-61198110\\
              Fax: +86-010-61198110\\
              \email{yusong@bupt.edu.cn}           %  \\
           \and
           Zhengyu Li \and Hong Guo\at
              State Key Laboratory of Advanced Optical Communication Systems and Networks, Center for Computational Science and Engineering and Center for Quantum Information Technology, School of Electronics Engineering and Computer Science, Peking University, Beijing 100871, China
}

\date{Received: date / Accepted: date}

\maketitle

\begin{abstract}
We propose a method to improve the performance of two-way continuous-variable quantum key distribution protocol by virtual photon subtraction. The Virtual photon subtraction implemented via non-Gaussian post-selection not only enhances the entanglement of two-mode squeezed vacuum state but also has advantages in simplifying physical operation and promoting efficiency. In two-way protocol, virtual photon subtraction could be applied on two sources independently. Numerical simulations show that the optimal performance of renovated two-way protocol is obtained with photon subtraction only used by Alice. The transmission distance and tolerable excess noise are improved by using the virtual photon subtraction with appropriate parameters. Moreover, the tolerable excess noise maintains a high value with the increase of distance so that the robustness of two-way continuous-variable quantum key distribution system is significantly improved, especially at long transmission distance.
\end{abstract}

\section{Introduction}
\label{intro}
Continuous-variable quantum key distribution~(CV-QKD) allows unconditional secure key distribution between two legal users Alice and Bob~\cite{CV,CV2}. CV-QKD protocol based on Gaussian modulated coherent state~\cite{GMCS1,GMCS2} has been rapidly developed in the past few years due to its natural advantages in combining with commercial telecom system~\cite{COMMERCIAL1,COMMERCIAL2}. As the technology of components and information reconciliation matures, the maximal transmission distance and the secure key rate of CV-QKD system have been continually promoted~\cite{EXPERIMENT1,EXPERIMENT2,EXPERIMENT3}. But the tolerable excess noise at long distance is still not enough to achieve a robust commercial CV-QKD system. Thus the improvement of tolerable excess noise is always an attractive field in CV-QKD research. Here, we propose a two-way CV-QKD protocol [10] enhanced by a round-trip quantum channel to improve robustness against excess noise. A more feasible two-way CV-QKD protocol, which is easier to implement and whose security can easier be analyzed, can be obtained by replacing Alice's displacement operation with a beam splitter~\cite{TWOBS,TWOAM}. Although the two-way protocol tolerates more excess noise than the one-way protocol, the level of tolerance decreases rapidly with distance.

Two-mode squeezed vacuum~(TMSV) state is used as the source of entanglement-based~(EB) CV-QKD protocol. It has been demonstrated that the entanglement of TMSV state is improved equivalently by implementing quantum operation in the prepare-and-measure~(PM) scheme such as the noiseless linear amplification~\cite{NLA} and thus to improve the performance of CV-QKD protocol~\cite{NLACV1,NLACV2,NLACV3}. In this paper, We propose a method to improve the performance of two-way CV-QKD protocol with virtual photon subtraction. Photon subtraction~\cite{PS1,PS2,PS3,PS4} is a kind of probabilistic quantum operation which has been used to enhance the entanglement of TMSV state~\cite{PSEB}. However, the practical photon subtraction cannot approach the expected performance because of the imperfect efficiency of photon number resolving~(PNR) detector~\cite{PNR} used in photon subtraction. Fortunately, we can use the virtual photon subtraction not only to simplify the physical scheme but also to promote the efficiency of PNR detector to 100\%~\cite{PSPS}.

This paper is organized as follows. In Sec.~\ref{sec:1}, we review the virtual photon subtraction achieved by non-Gaussian post-selection and propose the two-way CV-QKD protocol with virtual photon subtraction. In Sec.~\ref{sec:2}, we derive the secure key rate of the two-way CV-QKD protocol with virtual photon subtraction. In Sec.~\ref{sec:3}, the numerical simulation results of the performance are presented with different system schemes. Conclusions are given in Sec.~\ref{sec:4}.
\section{Improvement of two-way CV-QKD protocol with virtual photon subtraction}
\label{sec:1}

In this section, we first review the virtual photon subtraction implemented by non-Gaussian post-selection~\cite{PSPS} in CV-QKD protocol. Then we present the EB scheme of two-way CV-QKD protocol with virtual photon subtraction.

\subsection{Virtual photon subtraction in CV-QKD protocol}

\begin{figure}[t]
\centering
\includegraphics[width=1\textwidth]{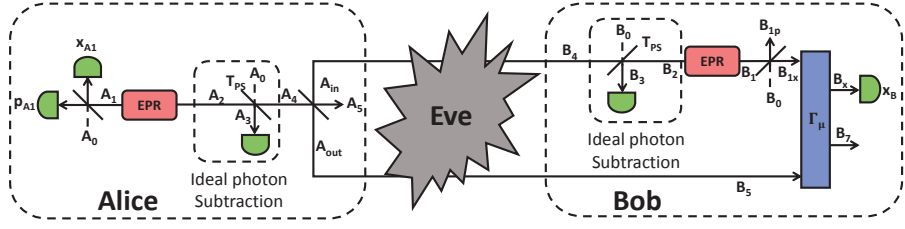}
\caption{Entanglement-based scheme of two-way CV-QKD protocol with virtual photon subtraction. Dash boxes represent the virtual photon subtraction used by Alice and Bob.}
\label{fig2}
\end{figure}

It has been confirmed that the virtual photon subtraction implemented via non-Gaussian post-selection is equivalent to the realistic photon subtraction with an ideal PNR detector~\cite{PSPS}. The EB scheme, which is more intuitive and easier to analyze, is shown in the dash box of Fig.~\ref{fig2}.

We assume that Alice applies virtual photon subtraction on her TMSV state, then the process of poset-selection is shown as follows. Alice first prepares a TMSV state and applies the heterodyne detection on mode ${A_1}$. A TMSV state is generated by squeezing a two mode vacuum state. The two mode squeezing operator is ${S_{TMS}}\left( r \right) = \exp \left[ {{{r\left( {{{\hat a}_1}{{\hat a}_2} - \hat a_1^\dag \hat a_2^\dag } \right)} \mathord{\left/
 {\vphantom {{r\left( {{{\hat a}_1}{{\hat a}_2} - \hat a_1^\dag \hat a_2^\dag } \right)} 2}} \right.
 \kern-\nulldelimiterspace} 2}} \right]$, where ${\hat a_1}$ and $\hat a_1^\dag $ ( ${\hat a_2}$ and $\hat a_2^\dag $ ) represent the annihilation and creation operator of modes ${A_1}$ and ${A_2}$ , $r$ is the squeezing parameter. Thus the TMSV state ${\left| \psi  \right\rangle _{{A_1}{A_2}}}$ can be described as

\begin{equation}
{\left| \psi  \right\rangle _{{A_1}{A_2}}} = {S_{TMS}}\left( r \right)\left| {0,0} \right\rangle  = \sqrt {1 - {\lambda ^2}} \sum\limits_{n = 0}^\infty  {{\lambda ^n}\left| {n,n} \right\rangle } ,
\end{equation}

Where $\lambda  = \tanh r$ , $\left| {n,n} \right\rangle  = {\left| n \right\rangle _{{A_1}}} \otimes {\left| n \right\rangle _{{A_2}}}$, $\left| n \right\rangle $ is Fock state.

There is an one-to-one correspondence between preparing a coherent state and measuring one mode of the TMSV state. When Alice obtains heterodyne detection results $\left( {{x_{A1}},{p_{A1}}} \right)$, the mode ${A_2}$ projects on $\left| \alpha  \right\rangle $, where $\alpha  = \sqrt 2 \lambda ({x_{A1}} - i{p_{A1}})$. The mode ${A_2}$ is separated into two modes ${A_3}$ and ${A_4}$ by a beam splitter (transmittance ${T_{PS}}$) and photon subtraction operation $M_{PS}^k = \left| k \right\rangle \left\langle k \right|$ is implemented on mode ${A_3}$. The photon subtraction $M_{PS}^k = \left| k \right\rangle \left\langle k \right|$ is a kind of measurement operation to confirm the photon number of quantum state, where $k$ is the photon number. It is implemented by the photon number resolving (PNR) detector. The mode ${A_4}$ is the output state of TMSV state with k photons subtraction and is described by

\begin{equation}
\rho _{{A_4}}^k = \int {\int {\frac{{P_{PS}^k\left( {k\left| {{x_{A1}},{p_{A1}}} \right.} \right)}}{{P_{PS}^k}}} } {P_{{x_{A1}}{p_{A1}}}}d{x_{A1}}d{p_{A1}}\left| {\sqrt {{T_{PS}}} \alpha } \right\rangle \left\langle {\sqrt {{T_{PS}}} \alpha } \right|,
\end{equation}

\begin{equation}
%\begin{aligned}
\begin{array}{l}
P_{PS}^k\left( {k\left| {{x_{A1}},{p_{A1}}} \right.} \right) = {\left| {\left\langle {k\left| {\sqrt {1 - {T_{PS}}} \alpha } \right\rangle } \right.} \right|^2},
\end{array}
%\end{aligned}
\end{equation}

Where ${T_{PS}}$ is transmittance of photon subtraction, $P_{PS}^k\left( {k\left| {{x_{A1}},{p_{A1}} }\right.} \right)$ is the success probability of k photons subtraction with Alice's heterodyne measurement results$\left( {{x_{A1}},{p_{A1}}} \right)$ and ${P_{{x_{A1}}{p_{A1}}}}$ is the two-dimension Gaussian distribution of Alice's measurement result $\left( {{x_{A1}},{p_{A1}}} \right)$.

In the case where the mode ${A_3}$ is not measured by $M_{PS}^k$ , the ${A_4}$ mode is described by

\begin{equation}
{\rho _{{A_4}}} = \int {\int {{P_{{x_{A1}}{p_{A1}}}}d{x_{A1}}d{p_{A1}}\left| {\sqrt {{T_{PS}}} \alpha } \right\rangle \left\langle {\sqrt {{T_{PS}}} \alpha } \right|} },
\end{equation}
By comparing ${\rho _{{A_4}}}$ with $\rho _{{A_4}}^k$ , it's easy to see we can get the photon subtracted TMSV state by selecting Alice¡¯s heterodyne measurement result $\left( {{x_{A1}},{p_{A1}}} \right)$ with a probability ${P_s}$

\begin{equation}
{P_s} = P_{PS}^k\left( {k\left| {{x_{A1}},{p_{A1}}} \right.} \right),
\end{equation}
Then the covariance matrix of the density matrix $\rho _{{A_1}{A_4}}^k$ representing the photon subtracted TMSV state, is given by

\begin{equation}
{\gamma _{{A_1}{A_4}}} = \left[ {\begin{array}{*{20}{c}}
{{V_{{A_1}}}{\rm I}}&{{C_{{A_1}{A_4}}}{\sigma _z}}\\
{{C_{{A_1}{A_4}}}{\sigma _z}}&{{V_{{A_4}}}{\rm I}}
\end{array}} \right],
\end{equation}
Where $\rm I$ is $\left[ {\begin{array}{*{20}{c}}
1&0\\
0&1
\end{array}} \right]$ and $\sigma _z$ is $\left[ {\begin{array}{*{20}{c}}
1&0\\
0&{ - 1}
\end{array}} \right]$.
Because of the symmetry of covariance matrix, we can get elements of the matrix by only calculating the variance and covariance of x quadrature

\begin{equation}\label{cm}
\begin{array}{l}
{V_{{A_1}}} = 2\int {\int {x_{A1}^2P\left( {{x_{A1}},{p_{A1}},{x_{A4}}} \right)d{x_{A1}}d{p_{A1}}d{x_{A4}}} }  - 1,\\
{C_{{A_1}{A_4}}} = \sqrt 2 \int {\int {{x_{A1}}{x_{A4}}P\left( {{x_{A1}},{p_{A1}},{x_{A4}}} \right)d{x_{A1}}d{p_{A1}}d{x_{A4}}} }, \\
{V_{{A_4}}} = \int {\int {x_{A4}^2P\left( {{x_{A1}},{p_{A1}},{x_{A4}}} \right)} } d{x_{A1}}d{p_{A1}}d{x_{A4}}.
\end{array}
\end{equation}
Where $P\left( {{x_{A1}},{p_{A1}},{x_{A4}}} \right) = \frac{{P_{PS}^k\left( {k\left| {{x_{A1}},{p_{A1}}} \right.} \right)}}{{P_{PS}^k}}{P_{{x_{A1}}{p_{A1}}}}{\left| {\left\langle {{x_{A4}}} \right.\left| {\sqrt {{T_{PS}}} \alpha } \right\rangle } \right|^2}$ ,\\$P_{PS}^k = \frac{{1 - {\lambda ^2}}}{{1 - {T_{PS}}{\lambda ^2}}}{\left[ {\frac{{{\lambda ^2}\left( {1 - {T_{PS}}} \right)}}{{1 - {T_{PS}}{\lambda ^2}}}} \right]^k}$. Obviously, the data after selection does not obey Gaussian distribution any more, then the original Gaussian state is transformed into  non-Gaussian state by above post-selection. Via post-selection, we can implement photon subtraction virtually as well as avoid the imperfect PNR detector and complex physical scheme.

\subsection{Two-way CV-QKD protocol with virtual photon subtraction}

Two-way CV-QKD protocol contains two sources located at Alice and Bob's sides respectively, the influence of implementing photon subtraction on two sources should be carefully analyzed. Since the function of two sources are different, the virtual photon subtraction implemented by Alice or Bob may generate different influence on the performance of two-way CV-QKD protocol. We will enhance the two-way CV-QKD protocol by the virtual photon subtraction in three different ways to obtain the following schemes. The two-way protocol with a beam splitter is used as the original two-way protocol~\cite{TWOBS}. The PM scheme of two-way CV-QKD protocol with virtual photon subtraction is described as follows and the equivalent EB scheme used in security analysis is shown in Fig.~\ref{fig2}:

Step1: Bob initially prepares a TMSV state with variance ${V_B}$ . Then he keeps one mode and sends another mode to Alice through the forward quantum channel.

Step2: Alice also prepares a TMSV state with variance ${V_A}$. She keeps one mode and couple another mode with the mode sent form Bob with a beam splitter (transmittance:~${T_A}$). Then Alice sends one of the output mode to Bob through the backward quantum channel and measures the another mode with homodyne detection. Eve can perform her attack both on the forward and backward quantum channel.

Step3: Alice and Bob measure the mode kept by themselves with heterodyne detection and get the
 variables $ ({x_A},{p_A}) $ and $ ({x_{B1}},{p_{B1}}) $ respectively. There is a one-to-one correspondence between preparing a coherent state and measuring one mode of a TMSV state.

Step4: Bob performs homodyne detection on the received state and get the variable ${x_{B5}}$ or ${p_{B5}}$. He uses ${x_B} = {x_{B5}} - \mu {x_{B1}}$ or ${p_B} = {p_{B5}} - \mu {p_{B1}}$ to estimate Alice's measurement result ${x_A}$ or ${p_A}$, where $\mu$ is the parameter used to optimize Bob's estimator.

Step5: Before the conventional post processing, Alice and Bob select a part of data with  probability $P_S^A$ and $P_S^B$ respectively(when Alice or Bob does not use subtracted source, $P_S^A$ or $P_S^B$ is set as \%100), and they reveals the selection result. Then Alice and Bob should keep the data which is selected by Bob and Alice simultaneously. By adjusting different selection probability $P_S^A$ and $P_S^B$ , Bob and Alice can achieve variable k photon(s) subtraction. Then Alice and Bob implement the reconciliation and privacy amplification through the classical channel.

To know the influence of a one-side photon subtraction on the other side's data, we first assume that the virtual photon subtraction via post-selection is only implemented by Alice. Alice selects her data $\left( {{x_A},{p_A}} \right)$ with a selection probability $P_S^A = P_{PS}^k\left( {k\left| {{x_A},{p_A}} \right.} \right)$ and the data of Bob is also selected with the probability $P_S^A$. Then the original Gaussian state of Alice will be transformed into a non-Gaussian state as equation~\ref{cm}. Since there is no correlation between Alice and Bob's data, the selection probability $P_S^A$ only derived by $\left( {{x_A},{p_A}} \right)$ is independent on Bob's data. As a consequence, when Bob uses the selection probability $P_S^A$ to select his data, it's just equivalent to select a group of Gaussian data randomly with the probability $P_S^A$ and the distribution of original TMSV state will not change. It means we can apply virtual photon subtraction on Alice or Bob's original TMSV state respectively without change of the other one. As a result, there are three two-way CV-QKD schemes with virtual photon subtraction: virtual photon subtraction only used by Alice, virtual photon subtraction only used by Bob, virtual photon subtraction used by both Alice and Bob.

Because a original Gaussian source state will be transformed into a non-Gaussian state via virtual photon subtraction, the secure key rate of two-way CV-QKD protocol with virtual photon subtraction ${K_{PS}}$ is no less than the key rate $K$ whose source is Gaussian state according to the optimality of Gaussian attack $\left( {{K_{PS}} \ge K} \right)$~\cite{GOP1,GOP2}. We can use the same covariance matrix of Gaussian state to estimate the lower bound of secure key rate with virtual photon subtraction.

When Alice and Bob use reverse reconciliation~\cite{REVERSE}, the secure key rate after virtual photon subtraction is

\begin{equation}\label{kr}
{K_{PS}} = P\left[ {\beta I\left( {B:A} \right) - S(E:B)} \right].
\end{equation}

Where $\beta $ is reconciliation efficiency, $P = P_{PS}^{{k_A}} \cdot P_{PS}^{{k_B}}$, $P_{PS}^{{k_A}}$ is the success probability of k photons subtracted source set at Alice's side as well as Bob, $I\left( {B:A} \right)$ is mutual information between Alice and Bob and $S(E:B) = S\left( E \right) - S\left( {E\left| B \right.} \right)$ is quantum mutual information between Eve and Bob. Because of the different function of Alice and Bob's sources, we should derive the secure key rate in three different cases respectively.

\section{Secure key rate of two-way CV-QKD protocol with photon subtraction}
\label{sec:2}
%\begin{figure}[t]
%\centering
%\includegraphics[width=6in]{Alice_PS.eps}
%\caption{Entanglement-based scheme of two-way CV-QKD when Alice uses photon subtracted source.
%}\label{fig4}
%\end{figure}

As introduced above, the PM scheme of virtual photon subtraction is equivalent to practical photon subtraction with a perfect PNR detector. So the EB scheme of two-way CV-QKD with virtual photon subtraction is shown in Fig.~\ref{fig2}. Then we use the EB scheme to derive security analysis. Although there are three EB schemes of two-way CV-QKD protocol with virtual photon subtraction, three methods of calculating the secure key rate are alike. To derive the secure key rate bound of the modified protocol, we just use the covariance matrix of photon subtracted source derived above and the process of virtual photon subtraction has been included. When virtual photon subtraction is used by Alice or Bob, the covariance matrix of their sources can be described by

\begin{equation}~\label{cmA}
{\gamma _{{A_1}{A_4}}} = \left[ {\begin{array}{*{20}{c}}
{{V_{{A_1}}}{\rm I}}&{{C_{{A_1}{A_4}}}{\sigma _z}}\\
{{C_{{A_1}{A_4}}}{\sigma _z}}&{{V_{{A_4}}}{\rm I}}
\end{array}} \right],
\end{equation}

\begin{equation}\label{cmB}
{\gamma _{{B_1}{B_4}}} = \left[ {\begin{array}{*{20}{c}}
{{V_{{B_1}}}{\rm I}}&{{C_{{B_1}{B_4}}}{\sigma _z}}\\
{{C_{{B_1}{B_4}}}{\sigma _z}}&{{V_{{B_4}}}{\rm I}}
\end{array}} \right],
\end{equation}

The elements of Alice's covariance matrix are derived by equation~\ref{cm}

\begin{equation}
\begin{array}{l}
{V_{{A_1}}} = 2{{V'}_A} - 1,\\
{C_{{A_1}{A_4}}} = 2\sqrt {{T_{PS}}} {\lambda _A}{{V'}_A},\\
{V_{{A_4}}} = 2{T_{PS}}\lambda _A^2{{V'}_A} + 1,\\
{{V'}_A} = \frac{{k + 1}}{{1 - {T_{PS}}\lambda _A^2}}.
\end{array}
\label{pscm}
\end{equation}

Elements of Bob's covariance matrix can be obtained just by changing A to B. According to above analysis, the one-side virtual photon subtraction will not change the other source. When Alice or Bob do not select their data, their source will remain in a Gaussian state. The corresponding covariance matrix can be derive by setting $k=0$ and ${T_{PS}} = 1$ in equation~\ref{pscm}.

Here, we assume Eve implements the one-mode entangling cloner attack~\cite{ATTACK} on two quantum channel independently. Thus the transmittance and excess noise of two channels are independent. Then the forward channel noise is ${\chi _1} = \frac{{(1 - {T_1})}}{{{T_1}}} + {\varepsilon _1}$ and the backward channel noise is ${\chi _2} = \frac{{(1 - {T_2})}}{{{T_2}}} + {\varepsilon _2}$.

After the transmission of quantum state, the initial system ${B_1}{B_4}{A_1}{A_4}$ is transformed into ${B_1}{B_5}{A_1}{A_5}E$. Since Eve can purify ${B_1}{B_5}{A_1}{A_5}E$, $S\left( E \right)$ can be calculated by $S\left( {{B_1}{B_5}{A_1}{A_5}} \right)$. The covariance matrix of state ${B_1}{B_5}{A_1}{A_5}$ is given by

\begin{equation}
{\gamma _{{B_1}{B_5}{A_1}{A_5}}} = \left[ {\begin{array}{*{20}{c}}
{{V_{{B_1}}}{\rm I}}&{{C_{{B_1}{B_5}}}{\sigma _Z}}&0&{{C_{{A_5}{B_1}}}{\sigma _Z}}\\
{{C_{{B_1}{B_5}}}{\sigma _Z}}&{{V_{{B_5}}}{\rm I}}&{{C_{{A_1}{B_5}}}{\sigma _Z}}&{{C_{{A_5}{B_5}}}{\sigma _Z}}\\
0&{{C_{{A_1}{B_5}}}{\sigma _Z}}&{{V_{{A_1}}}{\rm I}}&{{C_{{A_1}{A_5}}}{\sigma _Z}}\\
{{C_{{A_5}{B_1}}}{\sigma _Z}}&{{C_{{A_5}{B_5}}}{\sigma _Z}}&{{C_{{A_1}{A_5}}}{\sigma _Z}}&{{V_{{A_5}}}{\rm I}}
\end{array}} \right],
\end{equation}
Then Eve's entropy is given by

\begin{equation}
S\left( {{B_1}{B_5}{A_1}{A_5}} \right) = \sum\limits_{i = 1}^4 {G\left( {{\lambda _i}} \right)}.
\end{equation}
Where $G({\lambda _i}) = \frac{{{\lambda _i} + 1}}{2}{\log _2}\left( {\frac{{{\lambda _i} + 1}}{2}} \right) + \frac{{{\lambda _i} - 1}}{2}{\log _2}\left( {\frac{{{\lambda _i} - 1}}{2}} \right)$ , and ${\lambda _i}$ is the symplectic eigenvalue of covariance matrix ${\gamma _{{B_1}{B_3}{A_1}{A_5}}}$.

We suppose Bob applies homodyne measurement. He first uses ${x_B} = {x_{{B_5}}} - \mu {x_{{B_1}}}$ or ${p_B} = {p_{{B_5}}} - \mu {p_{{B_1}}}$ to estimate Alice's measurement results ${x_{{A_1}}}$ or ${x_{{A_1}}}$ which can be applied by a symplectic transformation ${\Gamma _\mu }$ in the EB scheme. In order to optimize Bob's estimation results, the parameter $\mu $ is set as $\mu  = \sqrt {2{T_A}{T_1}{T_2}{{\left( {{V_{{B_4}}} - 1} \right)} \mathord{\left/
{\vphantom {{\left( {{V_{B_4}} - 1} \right)} {\left( {{V_{B_1}} + 1} \right)}}} \right.
\kern-\nulldelimiterspace} {\left( {{V_{B_1}} + 1} \right)}}} $ here. Where ${\Gamma _\mu } = \left[ {\begin{array}{*{20}{c}}
1&{}&{ - \mu }&{}\\
{}&1&{}&{}\\
{}&{}&1&{}\\
{}&\mu &{}&1
\end{array}} \right]$ for $x$ and ${\Gamma _\mu } = \left[ {\begin{array}{*{20}{c}}
1&{}&{}&{}\\
{}&1&{}&\mu \\
{ - \mu }&{}&1&{}\\
{}&{}&{}&1,
\end{array}} \right]$ for $p$.

After transformation ${\Gamma _\mu }$, the matrix of ${B_{1p}}{B_x}{B_7}{A_1}{A_5}$ is
\begin{equation}
{\gamma _{{B_{1p}}{B_x}{B_7}{A_1}{A_5}}} = \left( {{\Gamma _\mu } \oplus {\rm I}} \right){\gamma _{{p_{B1}}{x_{B1}}{B_5}{A_1}{A_5}}}{\left( {{\Gamma _\mu } \oplus {\rm I}} \right)^T},
\end{equation}

Bob applies homodyne detection on ${B_x}$ resulting in ${x_B}$ quadrature and the classical mutual information is

\begin{equation}
I\left( {A:B} \right) = \frac{1}{2}\log \left( {{V_{{A_x}}}} \right) - \frac{1}{2}\log \left( {{V_{{A_x}\left| {{B_x}} \right.}}} \right) = \frac{1}{2}\log \left( {\frac{{{V_{{A_1}}} + 1}}{{{V_{{A_1}}} - \frac{{C_{{A_1}{B_x}}^2}}{{{V_{{B_x}}}}} + 1}}} \right),
\end{equation}
Where ${V_{{A_1}}}$ ,${V_{{B_x}}}$ , ${C_{{A_1}{B_x}}}$ are variance for mode ${A_1}$ , ${B_x}$ and their covariance respectively. Then condition covariance matrix $\gamma _{{B_{1p}}{B_7}{A_1}{A_5}}^{{x_{B}}}$ after Bob's measurement is

\begin{equation}
\gamma _{{B_{1p}}{B_7}{A_1}{A_5}}^{{x_{B}}} = {\gamma _{{B_{1p}}{B_7}{A_1}{A_5}}} - {C_{{B_x}}}{\left( {{X_x}{\gamma _{{B_x}}}{X_x}} \right)^{MP}}C_{{B_x}}^T,
\end{equation}
Where ${X_x} = {\rm{diag}}(1,0)$ and MP denotes the Moore-Penrose inverse of the matrix. Eve's condition entropy is  $S\left( {E\left| {{x_B^M}} \right.} \right)$. Since ${B_{1p}}{B_7}{A_1}{A_5}E$ is a pure state, then $S\left( {E\left| {{x_B^M}} \right.} \right) = S\left( {{B_{1p}}{B_7}{A_1}{A_5}\left| {{x_B^M}} \right.} \right)$. Eve's condition entropy is given by

\begin{equation}
S\left( {{B_{1p}}{B_7}{A_1}{A_5}\left| {{x_B^M}} \right.} \right) = \sum\limits_{i = 5}^8 {G\left( {{\lambda _i}} \right)},
\end{equation}
Where $\lambda $ is the symplectic eigenvalue of the condition covariance matrix
\\$\gamma _{{B_{1p}}{B_7}{A_1}{A_5}}^{{x_{B}}}$. Because of the independence of Alice and Bob's post-selection, the success probability of virtual photon subtraction is given by $P = P_{PS}^{{k_A}} \cdot P_{PS}^{{k_B}}$. When Alice or Bob doesn't use virtual photon subtraction, success probability $P_{PS}^{{k_A}}$ or $P_{PS}^{{k_B}}$ is 100\%.

By substituting above result into equation~\ref{kr}, the final secure key rate can be written by

\begin{equation}
{K_{PS}} = P\left[ {\beta I\left( {B:A} \right) - \sum\limits_{i = 1}^4 {G\left( {{\lambda _i}} \right)}  + \sum\limits_{i = 5}^8 {G\left( {{\lambda _i}} \right)} } \right].
\end{equation}

\section{Simulation and analysis}
\label{sec:3}
Since virtual photon subtraction can be used independently by Alice or Bob, we utilize the numerical simulation of the secure key rate and tolerable excess noise which are crucial performance of CV-QKD protocol to search the best scheme of three cases.

\subsection{Two-way CV-QKD protocol with photon subtraction only used by Alice}

\begin{figure}[t]
\centering
\includegraphics[width=1\textwidth]{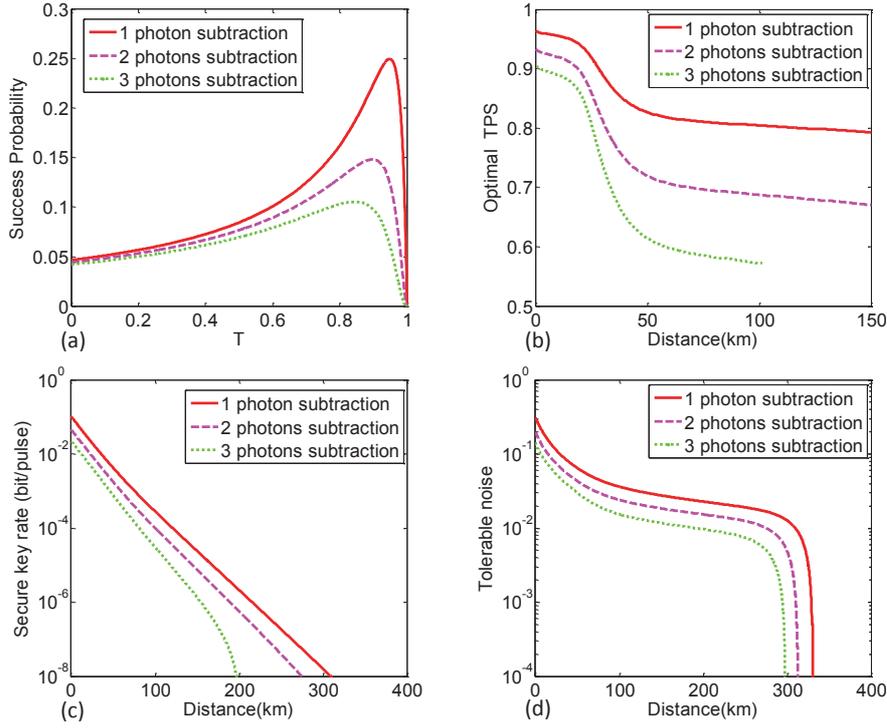}
\caption{(a) The success probability $P_{PS}^k$ of virtual photon subtraction by setting different transmittance ${T_{PS}}$. (b) The optimal ${T_{PS}}$ for the secure key rate when different k photon(s) subtraction is used by Alice. (c) The secure key rate with optimal ${T_{PS}}$ in (b) when different k photon(s) subtraction is used by Alice. (d) The tolerable excess noise when different k photon(s) subtraction is used by Alice. The variance of TMSV state is ${V_A} = {V_B} = 40$, the excess noise is $\varepsilon  = 0.01$, the reconciliation efficiency is $\beta  = 0.95$, the beam splitter transmittance of Alice ${T_A} = 0.5$.
}\label{SAlice1}
\end{figure}

In two-way CV-QKD protocol, the information of secure key is modulated on the source of Alice. As a result, replacing Alice's original Gaussian-modulated coherent state with photon subtracted source may lead to significant change of performance. Compared to the original two-way protocol, there are two additional parameters photon subtraction number ${k_A}$ and transmittance ${T_{PSA}}$ in the two-way CV-QKD protocol with virtual photon subtraction. At first, we show the success probability $P_{PS}^{{k_A}}$ of virtual photon subtraction by setting different transmittance ${T_{PSA}}$ in Fig.~\ref{SAlice1}(a) when the variance of original TMSV state is constant. When virtual photon subtraction succeeds, secure key rate ${K_S} = \beta I\left( {B:A} \right) - S(E:B)$ is also changed with different transmittance ${T_{PSA}}$. Thus there always is a tradeoff between the original secure key rate ${K_S}$ and the success probability $P_{PS}^{{k_A}}$ . We can find the optimal transmittance ${T_{PSA}}$ which is illustrated in Fig.~\ref{SAlice1}(b) to get the maximum final secure key rate ${K_{PS}}$ with varied transmission distance. Then, we consider the influence of different(1,2,3) photons subtraction on performance with above parameters. As shown in Fig.~\ref{SAlice1}(c), the modified protocol with 1 photon subtraction used by Alice get the highest secure key rate and the furthest maximal transmission distance (the secure key rate higher than 1e-8 is reserved). The tolerable excess noise of CV-QKD protocol under three situations is shown in Fig.~\ref{SAlice1}(d), the modified protocol with 1 photon subtraction used by Alice still performs better than other cases. The influence of different photon(s) subtraction used by Alice on performance coincides well with previous research result where 1 photon subtracted source is optimal in one-way CV-QKD protocol~\cite{PSPS}.

\begin{figure}[t]
\centering
\includegraphics[width=1\textwidth]{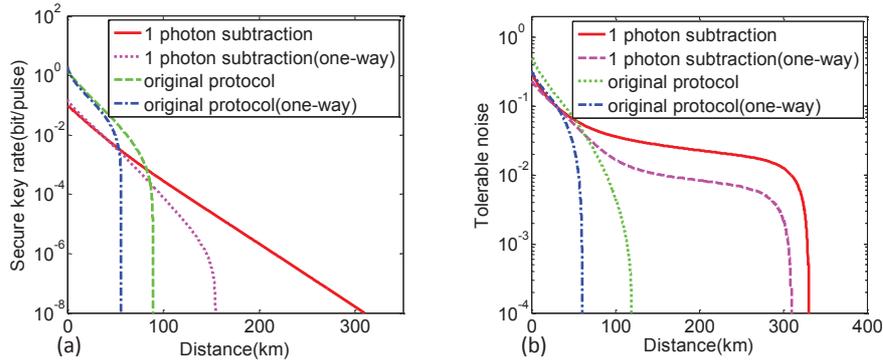}
\caption{(a) The comparison of secure key rate among the two-way CV-QKD protocol with 1 photon subtraction only used by Alice, the original two-way CV-QKD protocol and the one-way CV-QKD protocol with 1 photon subtraction. (b) The comparison of tolerable excess noise among the two-way CV-QKD protocol with 1 photon subtraction only used by Alice, the original two-way CV-QKD protocol and the one-way CV-QKD protocol with 1 photon subtraction. The variance of TMSV state is ${V_A} = {V_B} = 40$ for two-way protocol and $V = 40$  for one-way protocol, the excess noise is $\varepsilon  = 0.01$, the reconciliation efficiency is $\beta  = 0.95$, the beam splitter transmittance of Alice is ${T_A} = 0.5$.
}\label{SAlice2}
\end{figure}

Secondly, we consider the performance improvement of the modified protocol by comparing to the original two-way CV-QKD protocol, the original one-way CV-QKD protocol and the one-way CV-QKD protocol with 1 photon subtraction respectively. The one-way protocol used in this paper is GG02~\cite{GMCS1}(coherent state and homodyne detection). The parameters of two-way CV-QKD protocol remain unchanged and the variance of TMSV state is $V = 40$, the excess noise is $\varepsilon  = 0.01$, the reconciliation efficiency is $\beta  = 0.95$ for one-way protocol. Fig.~\ref{SAlice2}(a) indicates that the modified two-way protocol gets the farthest transmission distance with a significant increase and the highest secure key rate at long distance. As is well known, two-way scheme and virtual photon subtraction can both improve the transmission distance of CV-QKD protocol. When we use virtual photon subtraction in two-way scheme, we can get their advantage on transmission distance at the same time. We show the comparison of tolerable excess noise among these protocols in Fig.~\ref{SAlice2}(b). The tolerable excess noise of all protocols are in the same order of magnitude within a short transmission distance, but the modified two-way protocol performs better than others with the increase of distance. Moreover, the tolerable excess noise of the modified two-way protocol keeps a high value instead of continuous declination.

\subsection{Two-way CV-QKD protocol with photon subtraction only used by Bob}

\begin{figure}[t]
\centering
\includegraphics[width=1\textwidth]{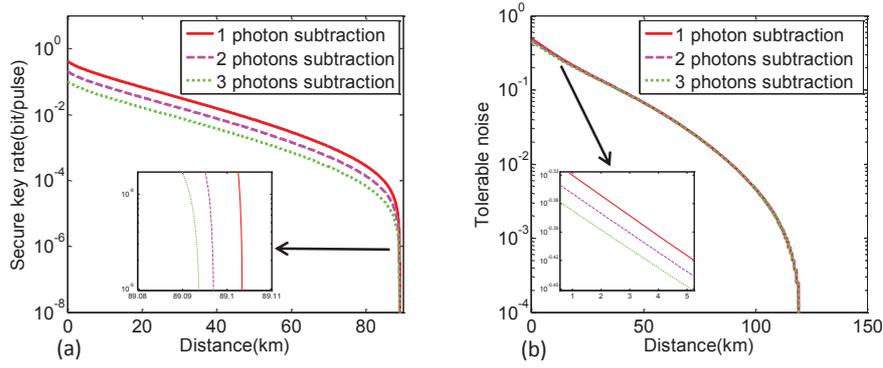}
\caption{(a) The secure key rate with optimal ${T_{PS}}$ when different k photon(s) subtraction is only used by Bob. (b) The tolerable excess noise when different k photon(s) subtraction is only used by Bob. The variance of TMSV state is ${V_A} = {V_B} = 40$, the excess noise is $\varepsilon  = 0.01$, the reconciliation efficiency is $\beta  = 0.95$, the beam splitter transmittance of Alice is ${T_A} = 0.5$.
}\label{SBob1}
\end{figure}

The source of Bob in two-way CV-QKD protocol is always used as auxiliary which doesn't contain the information of key rate. Hence, whether there is a virtual photon subtraction at Bob's side or not, we can see that his final estimator of Alice's data keeps almost unchanged when an optimal $\mu $ is used. Thus when different photon(s) subtraction is only used by Bob, the maximal transmission distance changes little as Fig.~\ref{SBob1}(a) shows. The difference of secure key rate is generated by the lower success probability of more photon(s) subtraction. Since the tolerable excess noise is independent on success
probability of virtual photon subtraction, the tolerable excess noise under three conditions is almost identical shown in Fig.~\ref{SBob1}(b).

\begin{figure}[t]
\centering
\includegraphics[width=1\textwidth]{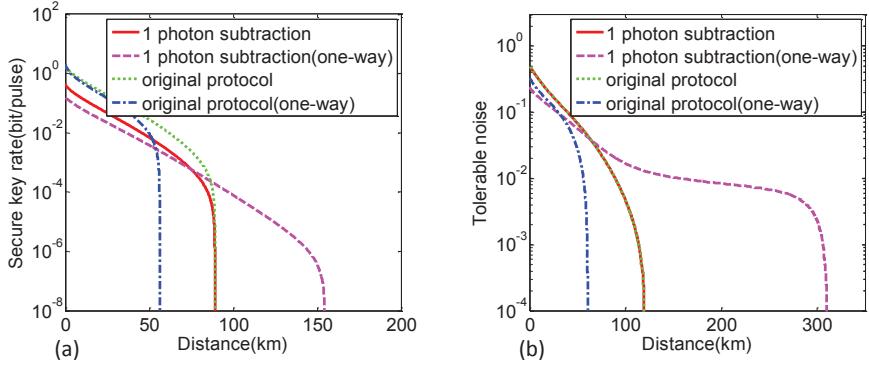}
\caption{The comparison of secure key rate among the two-way CV-QKD protocol when 1 photon subtraction is only used by Bob, the original two-way CV-QKD protocol and the one-way CV-QKD protocol with 1 photon subtraction. (b) The comparison of tolerable excess noise among the two-way CV-QKD protocol when 1 photon subtraction is only used by Bob, the original two-way CV-QKD protocol and the one-way CV-QKD protocol with 1 photon subtraction. The variance of TMSV state is ${V_A} = {V_B} = 40$ for two-way protocol and $V = 40$ for one-way protocol, the excess noise is $\varepsilon  = 0.01$, the reconciliation efficiency is $\beta  = 0.95$, the beam splitter transmittance of Alice is ${T_A} = 0.5$.
}\label{SBob2}
\end{figure}

We compare two-way CV-QKD protocol with 1 photon subtraction used by Bob to the original two-way CV-QKD protocol, and the one-way CV-QKD protocol with 1 photon subtraction respectively. As to the above analyses, the maximal transmission distance cannot be improved than the original two-way CV-QKD protocol which is shown in Fig.~\ref{SBob2}(a). Moreover, the secure key rate when Bob uses photon subtracted source is lower than the original protocol since the success probability cannot achieve 100\%. Because the tolerable excess noise is independent on the success probability of virtual photon subtraction, the tolerable excess noise of the modified two-way protocol is identical to the original two-way protocol shown in Fig.~\ref{SBob2}(b). The result implies two-way CV-QKD protocol with virtual photon subtraction only used by Bob can not improve the performance of protocol and the limited success probability which means lower use ratio of data even leads to the decrease of secure key rate.

\subsection{Two-way CV-QKD protocol with photon subtraction used by Alice and Bob}

\begin{figure}[t]
\centering
\includegraphics[width=1\textwidth]{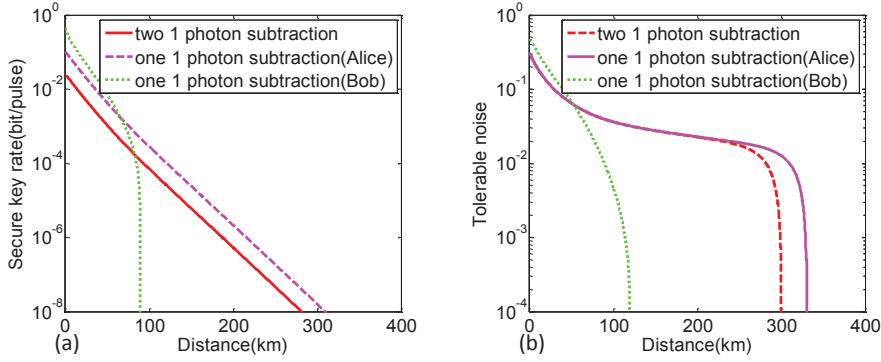}
\caption{(a) The comparison of secure key rate among the two-way CV-QKD protocol with two 1 photon subtraction, the original two-way CV-QKD protocol and the one-way CV-QKD protocol with 1 photon subtraction. (b) The comparison of tolerable excess noise among the two-way CV-QKD protocol with two 1 photon subtraction, the original two-way CV-QKD protocol and the one-way CV-QKD protocol with 1 photon subtraction. The variance of TMSV state is ${V_A} = {V_B} = 40$ for two-way protocol and $V = 40$ for one-way protocol, the excess noise is $\varepsilon  = 0.01$, the reconciliation efficiency is $\beta  = 0.95$, the beam splitter transmittance of Alice is ${T_A} = 0.5$.
}\label{SAliceBob}
\end{figure}

According to the results of two-way CV-QKD protocol with virtual photon subtraction used by Alice or Bob, the performance of system is only improved when virtual photon subtraction is only used by Alice and 1 photon subtraction performs better than more photons subtraction. Then we compare the performance of scheme with two virtual photon subtraction to the scheme with virtual photon subtraction is only by Alice or Bob alone. The comparison of secure key rate shown in Fig.~\ref{SAliceBob}(a) indicates we cannot enhance the secure key rate higher by using two virtual photon subtraction at each side than the scheme with virtual photon subtraction only used by Alice. For tolerable excess noise, there is a similar result shown in Fig.~\ref{SAliceBob}(b) where the tolerable excess noise of two-way CV-QKD protocol with two virtual photon subtraction is not better than the scheme with virtual photon subtraction only sued by Alice.

In two-way CV-QKD protocol, Alice's source (TMSV) is used to carry key information. So when virtual photon subtraction is used by Alice, the performance of two-way CV-QKD protocol is improved. But the information of Bob's source (TMSV) is nearly removed by a CNOT operation. Thus when Bob uses the virtual photon subtraction, the performance of two-way CV-QKD protocol should remain unchanged. Moreover, virtual photon subtraction is a probabilistic operation. The probability of failure leads to the decrease of the final performance of two-way CV-QKD protocol with virtual photon subtraction used by Bob. Two sources of two-way CV-QKD protocol are independent, the enhancement effect of Alice¡¯s photon subtracted source and the reduction effect of Bob¡¯s photon subtracted source will affect the performance simultaneously when two virtual photon subtraction are used. Thus the optimal scheme is virtual photon subtraction only used by Alice.

\section{Conclusion}
\label{sec:4}
In this paper, we propose a method to improve the performance of two-way continuous-variable quantum key distribution with virtual photon subtraction by enhancing the entanglement of TMSV state. Virtual photon subtraction can be applied on two sources of two-way CV-QKD protocol independently, thus there are three two-way CV-QKD schemes with virtual photon subtraction: virtual photon subtraction only used by Alice, virtual photon subtraction only used by Bob, virtual photon subtraction used by both Alice and Bob. Numerical simulations show that the optimal scheme of modified protocol is virtual photon subtraction only used by Alice and the performance of two-way CV-QKD protocol is improved by adjusting appropriate parameters of virtual photon subtraction. Furthermore, this method provides stable and high tolerable excess noise which is able to support a robust CV-QKD system at long distance over $200$km.

\section*{Acknowledgments}

This work was supported in part by the National Basic Research Program of China (973 Pro-gram) under Grant 2014CB340102, in part by the National Natural Science Foundation under Grants 61225003, 61531003, 61427813, 61401036, 61471051.


\begin{thebibliography}{}
\bibitem{CV}
Weedbrook C, Pirandola S, Garc\'ia-Patr\'on~R, Cerf~N~J, Ralph~T~C, Shapiro J H, and Lloyd S. Gaussian quantum information. Rev. Mod. Phys. 84(2): 621(2012)

\bibitem{CV2}
Wang X B, Hiroshima T, Tomita A, and Hayashi M. Quantum information with Gaussian states. Phys. Rep. 448(1): 1-111(2007)
%A. M. Lance, T. Symul, V. Sharma, C. Weedbrook, T. C. Ralph, and P. K. Lam, Phys. Rev. Lett. 95, (2005) 180503.

\bibitem{GMCS1}
Grosshans F, Grangier P. Continuous variable quantum cryptography using coherent states. Phys. Rev. Lett. 88(5): 057902(2002)

\bibitem{GMCS2}
Weedbrook~C, Lance~A~M, Bowen~W~P, Symul~T, Ralph~T~C and Lam~P~K. Quantum cryptography without switching. Phys. Rev. Lett. 93(17): 170504(2004)

\bibitem{COMMERCIAL1}
Qi B, Zhu W, Qian L and Lo H K. Feasibility of quantum key distribution through a dense wavelength division multiplexing network. New J. Phys. 12(10): 103042(2010)

\bibitem{COMMERCIAL2}
Kumar R, Qin H and All¨¦aume R. Coexistence of continuous variable QKD with intense DWDM classical channels. New J. Phys. 17(4): 043027(2015)

\bibitem{EXPERIMENT1}
Lodewyck~J, Bloch~M, Garc\'ia-Patr\'on~R, Fossier~S, Karpov~E, Diamanti~E, Debuisschert~T, Cerf~N~J, Tualle-Brouri~R, McLaughlin~S~W and Grangier~P. Quantum key distribution over 25 km with an all-fiber continuous-variable system. Phys. Rev. A. 76(4): 042305(2007)

\bibitem{EXPERIMENT2}
Jouguet~P, Kunz-Jacques~S, Leverrier~A, Grangier~P and Diamanti~E. Experimental demonstration of long-distance continuous-variable quantum key distribution. Nat. Photon. 7(5): 378-381(2013)

\bibitem{EXPERIMENT3}
Huang D, Lin D, Wang C, Liu W, Fang S, Peng J, Huang P and Zeng G. Continuous-variable quantum key distribution with 1 Mbps secure key rate. Opt. Express. 23(13): 17511-17519(2015)

%\bibitem{GPS}
%Walk N, Ralph T C, Symul T and Lam P K 2013 \emph{Phys. Rev. A} \textbf{87} 020303
%
%\bibitem{TWOA}
%Zhang Y C, Li Z, Weedbrook C, Yu S, Gu W, Sun M, Peng X and Guo H 2014 \JPB \textbf{47} 035501

\bibitem{TWO}
Pirandola~S, Mancini~S, Lloyd~S and Braunstein~S~L. Continuous-variable quantum cryptography using two-way quantum communication. Nat. Phys.~4(9): 726-730(2008)

\bibitem{TWOBS}
Sun~M, Peng~X, Shen~Y and Guo~H. Security of a new two-way continuous-variable quantum key distribution protocol. Int. J. Quantum Inform. 10(05): 1250059(2012)

\bibitem{TWOAM}
Zhang Y, Li Z, Weedbrook C, Yu S, Gu W, Sun M, Peng X and Guo H. Improvement of two-way continuous-variable quantum key distribution using optical amplifiers. J. Phys. B. 47(3): 035501(2014)

\bibitem{NLA}
Xiang G Y, Ralph T C, Lund A P, Walk N and Pryde G J. Heralded noiseless linear amplification and distillation of entanglement. Nat. Phys. 4(5): 316-319(2010)

\bibitem{NLACV1}
Blandino R, Leverrier A, Barbieri M, Etesse J, Grangier P and Tualle-Brouri R. Improving the maximum transmission distance of continuous-variable quantum key distribution using a noiseless amplifier. Phys. Rev. A. 86(1): 012327(2012)

\bibitem{NLACV2}
Zhang Y, Li Z, Weedbrook C, Marshall K, Pirandola S, Yu S and Guo H. Noiseless linear amplifiers in entanglement-based continuous-variable quantum key distribution. Entropy. 17(7): 4547-4562(2015)

\bibitem{NLACV3}
Zhang Y, Yu S and Guo H. Application of practical noiseless linear amplifier in no-switching continuous-variable quantum cryptography. Quantum Inf Process. 14(11): 4339-4349(2015)

\bibitem{PS1}
Opatrn\'y T, Kurizki G and Welsch D G. Improvement on teleportation of continuous variables by photon subtraction via conditional measurement. Phys. Rev. A. 61(3): 032302(2000)

\bibitem{PS2}
M S Kim, E Park, Knight P L , and Jeong H. Nonclassicality of a photon-subtracted Gaussian field. Phys. Rev. A. 71(4): 043805(2005)

\bibitem{PS3}
Kitagawa A, Takeoka M, Sasaki M and Chefles A. Entanglement evaluation of non-Gaussian states generated by photon subtraction from squeezed states. Phys. Rev. A. 73(4): 042310(2006)

\bibitem{PS4}
Navarrete-Benlloch C, Garc\'ia-Patr\'on~R, Shapiro J H and Cerf N J. Enhancing quantum entanglement by photon addition and subtraction. Phys. Rev. A. 86(1): 012328(2012)

\bibitem{PSEB}
Huang P, He G, Fang J and Zeng G. Performance improvement of continuous-variable quantum key distribution via photon subtraction. Phys. Rev. A. 87(1): 012317(2013)

\bibitem{PNR}
Eisaman M D, Fan J, Migdall A and Polyakov S V. Invited review article: Single-photon sources and detectors[J]. Rev. Sci. Instrum. 82(7): 071101(2011)

\bibitem{PSPS}
Li Z, Zhang Y C, Wang X, Xu B, Peng X and Guo H. Non-Gaussian postselection and virtual photon subtraction in continuous-variable quantum key distribution. Phys. Rev. A. 93(1): 012310(2016)

\bibitem{GOP1}
Navascu\'es M, Grosshans F and Ac\'in A. Optimality of Gaussian attacks in continuous-variable quantum cryptography. Phys. Rev. Lett. 97(19): 190502(2006)

\bibitem{GOP2}
Garc\'ia-Patr\'on~R and Cerf N J. Unconditional optimality of Gaussian attacks against continuous-variable quantum key distribution. Phys. Rev. Lett. 97(19): 190503(2006)

\bibitem{REVERSE}
Grosshans~F, Van Assche~G, Wenger~J, Brouri~R, Cerf~N~J and Grangier~P. Quantum key distribution using gaussian-modulated coherent states. Nature. 421(6920): 238-241(2003)

\bibitem{ATTACK}
Weedbrook C, Grosse N B, Symul T, Lam P K and Ralph T C. Quantum cloning of continuous-variable entangled states. Phys. Rev. A. 77(5): 052313(2008)
% etc
\end{thebibliography}
\end{document}